\documentclass[11pt,english]{article}
\pdfoutput=1
\usepackage{jheppub}
\usepackage{lscape}
\usepackage{amsmath}
\usepackage{slashed}
\usepackage{mathtools}
\usepackage{tikz-cd}
\usetikzlibrary{shapes,arrows,shadows}
\usepackage{amsmath,bm,times}
\usepackage{cleveref}
\usepackage{tcolorbox}
\usepackage[color=teal!40]{todonotes}
\usepackage{diagbox}
\makeatletter
\newcommand{\@fpheader}{}
\makeatother

\usepackage[english]{babel}

\usepackage{amsfonts}
\usepackage{amssymb}
\usepackage{amsmath}
\usepackage{graphicx}

\begin{document}

\title{Double fibration in G-theory and the cobordism conjecture}


\author[a]{Cesar Damian,} 
\author[b]{Oscar Loaiza-Brito,}
\author[b]{V\'ictor M. L\'opez-Ramos}

\affiliation[a]{Departamento de Ingenier\'ia Mec\'anica, Universidad de Guanajuato, \\
Carretera Salamanca-Valle de Santiago km 3.5+1.8 Comunidad de Palo Blanco, Salamanca, Mexico}
\affiliation[b]{Departamento de F\'isica, Universidad de Guanajuato, \\
Loma del Bosque No. 103 Col. Lomas del Campestre C.P 37150 Leon, Guanajuato, Mexico.}

\emailAdd{cesar.damian@ugto.mx,oloaiza@fisica.ugto.mx, vm.lopezramos@ugto.mx}

\date{\today}
\today
\abstract
{We investigate Type IIB compactifications with spatially varying fluxes and dilaton profiles in the setting of dynamical cobordism. In particular, we analyze a G-theory–motivated compactification in which the fluxes and the dilaton depend on coordinates of a complex two-dimensional plane. From the equations of motion, we deduce the existence of End-of-the-World branes. In a cohomological interpretation, these branes appear precisely in order to trivialize the relevant cohomology class. Furthermore, we compute the associated bordism group and show that additional non-perturbative objects are needed to cancel the class, while retaining the cohomological contribution as a subgroup. This suggests a mathematical structure that connects energy scales with the emergence of perturbative and non-perturbative physics.
}

\maketitle

\section{Introduction}
The absence of global symmetries in quantum gravity \cite{Banks:2010zn, Gaiotto:2014kfa, Harlow:2018tng} (see also \cite{Sharpe:2015mja, Benedetti:2022zbb, Cordova:2022ruw, Gomes:2023ahz, Brennan:2023mmt, Bhardwaj:2023kri, Luo:2023ive, Loaiza-Brito:2024hvx}) has emerged as a cornerstone principle within the swampland program~\cite{Palti:2019pca, vanBeest:2021lhn, Grana:2021zvf, Agmon:2022thq, Makridou:2023wkb}. Extending this idea to include higher-form symmetries has led to the Cobordism Conjecture \cite{McNamara:2019rup} (see \cite{Montero:2020icj, Blumenhagen:2021nmi, Blumenhagen:2022bvh} for some progress), which posits that the total cobordism group in a consistent theory of quantum gravity must be trivial. In physical terms, any would-be global topological charge must be either gauged or broken by the inclusion of suitable extended objects, ensuring that no exact global remnant remains. This framework predicts the existence of new dynamical ingredients, such as end-of-the-world (ETW) branes, orientifolds, or other exotic defects that cancel cobordism charges and terminate spacetime consistently.\\

A dynamical realization of this principle is provided by the Dynamical Cobordism Conjecture \cite{burattiDynamicalCobordismSwampland2021,burattiDynamicaltadpolesstringy2021,blumenhagenDynamicalcobordismdomain2022}, which states that any solution to an effective supergravity theory developing a singularity at a finite proper distance must be completed by an explicit ETW brane that caps off spacetime. In this context, the presence of a dynamical tadpole, a violation of the dilaton or flux equations of motion characterized by an order parameter, acts as a source driving the evolution of moduli fields until a finite-distance boundary is reached. The backreaction of these components generates a non-trivial cobordism class, which is then trivialized by introducing appropriate defects that either gauge or break the corresponding global charge. The resulting process naturally leads to the dynamical compactification of some spacetime directions.\\

This phenomenon has been explicitly realized in several string constructions. In Type IIB string theory, the backreaction of RR and NSNS fluxes can cancel a dilaton tadpole, leading to a singular but physically consistent configuration where the spacetime ends on O8-planes or D8-branes codimension-two ETW branes that resolve the singular geometry \cite{dudasBranesolutionsstrings2000,blumenhagenDilatontadpoleswarped2001}. Similar mechanisms appear in well-known compactifications such as the Klebanov–Tseytlin \cite{klebanovGravitydualssupersymmetric2000} and Klebanov–Strassler \cite{klebanovSupergravityconfininggauge2000} solutions, where the inclusion of fractional branes and fluxes resolves apparent singularities through geometric transitions. The general picture that emerges from these studies is that a non-trivial cobordism class in an effective theory corresponds to the presence of a dynamical tadpole, and the consistent completion of the background requires additional extended objects whose inclusion trivializes the cobordism group.\\

A prototypical realization of this idea is given in \cite{blumenhagenDynamicalcobordismdomain2022}, where the backreaction of a neutral, non-supersymmetric D8-brane induces a finite-length geometry that must terminate on an ETW 7-brane. The resulting configuration provides an explicit example of Local Dynamical Cobordism, in which spacetime dynamically ends on a codimension-two defect. The finite proper length of the interval, the logarithmic behavior of the dilaton near the boundary, and the existence of a positive-tension brane all support the interpretation of the ETW brane as the physical realization of the cobordism boundary required by quantum gravity.\\

In parallel, higher-dimensional geometric frameworks such as F-theory and its extensions provide natural environments in which these cobordism phenomena can be studied. In F-theory, Type IIB backgrounds with varying axio-dilaton are geometrized via an auxiliary elliptic fibration, whose degenerations encode the presence of 7-branes and their associated monodromies~\cite{vafaEvidenceFtheory1996,morrisonCompactificationsFtheoryCalabiYau1996}. More generally, G-theory introduced in \cite{candelasTypeIIBflux2015,candelasTypeIIBflux2015a} extends this idea by promoting the elliptic fiber to a K3 surface, thus geometrizing the full U-duality group of Type II string theory. In this construction, the complex structure of the K3 fiber encodes not only the axio-dilaton but also additional flux degrees of freedom, while the degeneration loci correspond to the locations of various $(p,q)$-branes, orientifolds, and non-geometric defects.\\

The compactification space of G-theory, locally of the form $\mathcal{M}_4 \times \mathbb{C} \times T^2 \times T^2$, is defined by up to five holomorphic functions that describe the variation of the K3 complex structure over the base \cite{candelasTypeIIBflux2015a}. Since the internal manifold is six-dimensional, it is topologically cobordant to a point, implying the absence of any topological obstruction to decay. However, dynamical stability can still arise from physical constraints such as the Dominant Energy Condition, which may prevent the decay of certain compactifications by contributing poles in the effective stress-energy tensor. In this sense, the auxiliary K3 geometry in G-theory provides a higher-dimensional setting where cobordism trivialization and energetic protection can coexist.\\

An illustrative comparison can be drawn with the example proposed in \cite{etxebarriaNothingcertainstring2020}, where the authors analyzed the trivial cobordism group $\Omega_3^{\mathrm{Spin}}=0$ for three-dimensional manifolds. Considering a compact space $T^3 = T^2 \times S^1$, one can construct a four-manifold $B_4 = T^2 \times \mathcal{D}$ whose boundary is $T^3$, provided that the spin structure of $T^3$ extends to $B_4$. In F-theory, this corresponds to an elliptic fibration where the base disk $\mathcal{D}$ is pinched at points where the fiber degenerates, representing the loci of 7-branes. By analogy, in G-theory, compactifications that include $(p,q)$-branes lead to local geometries of the form $T^2 \times T^2 \times S^2$, and filling the $S^2$ naturally suggests a bounding manifold $B_7 = T^2 \times T^2 \times \mathcal{B}^3$, where $\mathcal{B}^3$ is a three-ball. This analogy hints that the auxiliary K3 fibration in G-theory plays the role of a higher-dimensional cobordism completion for the physical spacetime, in the same way that F-theory encodes cobordism trivialization via its elliptic fiber.\\

The purpose of this work is to explore how G-theory compactifications generalize the framework of dynamical cobordism, providing a geometric dictionary between cobordism charges and the topological invariants of the K3-fibered auxiliary space. By studying how ETW branes and flux-induced compactifications appear as smooth geometric transitions in the K3 fibration, we aim to establish a correspondence between dynamical cobordism and the global geometry of G-theory, offering a unified higher-dimensional realization of cobordism trivialization in string theory.\\

Specifically, we observe that, at the cohomological level, namely, starting from the equations of motion, we can derive constraints on the fluxes and warping factors by demanding the cancellation of global charges or, equivalently, by requiring that the equations of motion define trivial cohomology classes. In our concrete setup, an elliptic fibration of a 4-dimensional torus over the punctured sphere, the consistency conditions familiar from G-theory, such as the requirement of having 24 extended branes, are recovered through the computation of the relevant cohomology group. Furthermore, by promoting this analysis from cohomology to the appropriate cobordism group and invoking the cobordism conjecture, we conclude that non-perturbative input is necessary to fully cancel the cobordism charge.  We also provide mathematical and physical arguments that narrow the cobordism group to an essentially unique candidate. Once the order of the 2-primary part is fixed, an exponent bound, which we argue holds in the degrees relevant to the computation, leaves three abelian candidates; excluding the one that would lose the $\mathbb{Z}_4$ structure of $\mathrm{SL}(2,\mathbb{Z})$ narrows these to two, and explicit generating manifolds single out $\mathbb{Z}_4\oplus(\mathbb{Z}_2)^3$. A $\mathbb{Z}_4$-valued $\eta$-invariant computation, which we do not carry out here, is required to fix the order of these generators, and hence both to exclude $(\mathbb{Z}_2)^5$ and to select $\mathbb{Z}_4\oplus(\mathbb{Z}_2)^3$ over $\mathbb{Z}_4\oplus\mathbb{Z}_4\oplus\mathbb{Z}_2$. We further suggest that the mathematical refinement from cohomology to cobordism
has a physical analog involving not only access to higher energies but also the
inclusion of genuinely non-perturbative physics.\\

The structure of our work is as follows. In Section \ref{sec:gkp-warped}, we introduce the method used to identify trivial cohomology classes from the equations of motion within the well-known warped compactification constructed by Giddings, Kachru, and Polchinski. From this viewpoint, we recover the relation between the warp factor and the RR five-form flux, together with the imaginary self-duality condition on $G_3$. In Section \ref{sec:a-double-fibration}, we describe the double elliptic fibration scenario studied in G-theory, emphasizing the global consistency conditions it imposes. We then demonstrate that, analogously to the GKP warped compactification, these conditions can be derived by requiring the cancellation of the cohomological charge obtained directly from the equations of motion and Bianchi identities. Next, still in Section \ref{sec:cohomology-to-cobordism}, we analyze the same setup from a topological perspective by computing the relevant homology and bordism groups. Finally, in Section \ref{sec:conclusions}, we offer our concluding remarks. The appendices contain supplementary material, including detailed explicit computations that complete our exposition.

\section{GKP warped compactification and the global charge conjecture}
\label{sec:gkp-warped}

The conditions on fluxes and sources in a warped compactification of Type IIB theory found by Giddings, Kachru and Polchinski (GKP) \cite{Giddings:2001yu} can also be obtained from the perspective of generalized symmetries and the cobordism conjecture. The Type IIB action in the Einstein frame reads:
\begin{equation} \label{eq:actionIIB}
S_{\text{IIB}} = \frac{1}{2\kappa_{10}^2} \int \left[ R\ast_{10}1  - \frac{1}{2} \frac{d\tau \wedge \ast_{10} d\bar{\tau}}{(\text{Im} \, \tau)^2} - \frac{1}{2} \frac{G_3 \wedge \ast_{10} \bar{G}_3}{\text{Im} \, \tau} - \frac{1}{4} F_5 \wedge \ast_{10} F_5 - \frac{i}{4} \frac{C_4 \wedge G_3 \wedge \bar{G}_3}{\text{Im}\, \tau} \right]
\end{equation}

where $\tau=C_0+ie^{-\phi}$ and $G_3=F_3-\tau H_3$. The equations of motion for all fields are

\begin{eqnarray} 
d\!\left( *_{10}\, d\phi \right)
&=& e^{2\phi}\,F_{1} \wedge \ast_{10} F_{1}
-\frac{1}{2}\,e^{-\phi}\,H_{3} \wedge \ast_{10}H_{3}
+\frac{1}{2}\,e^{\phi}\,F_{3} \wedge \ast_{10}F_{3},
 \label{dilaton}\\
d F_5 &=& H_3 \wedge F_3 = \frac{i}{2e^{-\phi}} G_3 \wedge \bar{G}_3.\\
d G_3&=& F_1\wedge F_3.
\end{eqnarray}
The trivial solution, i.e., with all fluxes turned off, leads to a compactification on a Calabi-Yau manifold. By considering the usual 10-dimensional space-time as $M_4\times M_6$ with a
warped metric 
\begin{equation} \label{eq:metric}
ds_{10}^2 = e^{2A(y)} g_{\mu\nu}(x) \, dx^\mu dx^\nu + e^{-2A(y)} h_{mn}(y) \, dy^m dy^n,
\end{equation}
we can calculate the exterior components of the Ricci tensor, which involve the warped factor $A(y)$ and the energy-tensor contributions from the fluxes $G_3$ and $F_5=(1+\ast_{10})d\alpha\wedge dx^0\wedge dx^1\wedge dx^2 \wedge dx^3$. Einstein's equation in four dimensions is then written as

\begin{equation}
d\ast_6d\,e^{4A}=\frac{e^{4A}}{2 e^{-\phi}}G_3\wedge\ast_6\bar{G}_3+e^{-4A}d\alpha\wedge\ast_6 d\alpha,
\end{equation}
where we have taken  $R_{\mu\nu}(g)=0$. Notice that the 5-form $\ast_6 de^{4A}$ is not closed unless the fluxes $G_3$ and $F_5$ are null, implying that
\begin{equation}
    [\ast_6 de^{4A}]\notin \text{H}^5(M_6;\mathbb{Z}).
\end{equation}
Therefore, a non-zero global charge can be constructed as
\begin{equation}
    Q_A=\int_{M_6}d\ast_6 de^{4A},
\end{equation}
\begin{equation}
d\ast_6 d\alpha=-\frac{ie^{4A}}{2e^{-\phi}}G_3\wedge\bar{G}_3+e^{-4A}de^{4A}\wedge\ast_6d\alpha.
\end{equation}

Similarly to $\ast_6 de^{4A}$, the 5-form $\ast_6 d\alpha$ is not closed; hence, all the observations about $\ast_6 de^{4A}$ are also valid for $\ast_6 d\alpha$.\\

However, we see from these expressions that the form
\begin{equation}
    \ast_6 J_1=\ast_6d e^{4A}-\ast_6d\alpha + d\ast_6f_2,
\end{equation}
is globally exact by taking
$\alpha= e^{4A}$ and the imaginary self dual (ISD) condition $\ast_6 G_3=i G_3$, implying the conservation of the current $\ast_6 J_1$, i.e., $[\ast_6 J_1]=[0]\in \text{H}^5(M_6, \mathbb{Z})$  with a zero global charge since the current has been gauged.\\

There are many implications of this, which make the GKP solution such an extraordinary proposal. Let us review the implications. First of all, the dilaton equation can be expressed in terms of the 6-dimensional Hodge dual as 
\begin{equation}
    d\ast_6 d\phi=4dA\wedge\ast_6 d\phi+\frac{1}{2}\left(e^\phi F_3\wedge\ast_6 F_3-e^{-\phi}H_3\wedge\ast_6 H_3\right),
\end{equation}
where only the presence of constant internal 3-form fluxes has been considered. The ISD condition on $G_3$ implies that
\begin{equation}
    e^{-\phi}F_3\wedge\ast_6 F_3= e^{\phi}H_3\wedge \ast_6 H_3,
\end{equation}
allowing us to fix the dilaton to a constant value in all dimensions and defining the string coupling constant $g_s=e^{-2\phi}$. By assuming a supersymmetric solution, we have taken $C_0=0$, or equivalently, a scenario with no $D7$-branes\footnote{This implies that $dG_3=0$, i.e., $G_3\in\text{H}^3(M_6,\mathbb{Z})$, which corresponds to a supersymmetric solution.}. Hence, there is no dynamical tadpole or NS-tadpole\footnote{Notice that there are four scalar fields, $\phi, A(y), C_0$ and $\alpha$, which might contribute to an exact 6-form. Since we have no $D7$-branes and we are considering a SUSY solution, only $A$ and $\alpha$ contribute.}\\

Second, gauging the current $\ast_6 J_1$ allows for the presence of sources for the 4-form $\ast_6 f_2$. However, the contributions from the sources to Einstein's and Bianchi equations must cancel each other. As shown by GKP, the sources are 3-dimensional BPS objects, such as D-branes and orientifold planes, which precisely satisfy this constraint. From the point of view of the Non-Global Charge (NGC) conjecture, this is also necessary to make $F_5$ a closed form, since in the presence of these sources, 
\begin{equation}
    dF_5=H_3\wedge F_3 + 2\kappa_{10} T_3 \rho_3
\end{equation}
with $\rho_3=e^{4A}\sum_m\delta(y^m-y^m_0)\ast_6 1$, such that $F_5\in \text{H}^5(M_6,\mathbb{Z})$, or equivalently that
\begin{equation}
    \int_{M_6} dF_5 =0,
\end{equation}
canceling the topological tadpole. Therefore, the sources for $F_5$ are necessary for the current $\ast_6 J_1$ to be gauged, $\ast_6 j_1= dF_5$. Hence, from the point of view of the 0-symmetry in $M_6$ related to the sources, it is gauged, while from the point of view of the 4-form symmetry related to $F_5$, it is broken. This solution is in agreement with the NGC conjecture.\\

Finally, notice that the presence of a 5-form $F_5$, which relates to the warping factor, is essential. This solution requires $F_5$ to be a non-zero form. Also, the Bianchi identities for $H_3$ and $F_3$ are imposed; i.e., we are considering constant fluxes on $M_6$, a proposal in agreement with the NGC conjecture that implies the absence of 5-branes. Notice also that a non-constant flux can be considered as far as
\begin{equation}
    \int_{\Sigma^4}dH_3=0,
\end{equation}
for some $\Sigma^4$ and similarly for $F_3$. \\

\subsection{Dynamical cobordism in a toroidal IIB compactification}
The reason we could justify the non presence of global charges in the above scenario comes precisely from the fact that we are considering a compact manifold $M_6$. Therefore, we can pose the inverse question. Is it possible to induce a compactification by imposing the conditions to have a zero global charge?\\

A particular example presented in \cite{burattiDynamicaltadpolesstringy2021} describes a flux compactification on an isotropic 6-dimensional torus. In this context, the inclusion of ISD fluxes drives a spontaneous compactification in some directions, which can be understood by the fact that each of the forms $\ast_6 d\alpha$ and $\ast_6 d e^{4A}$ is separately non-closed. The idea is to start with a non-compact internal space of the form $T^5\times \mathbb{R}$, where the extended direction is parameterized by the coordinate $y$. There is a constant ISD flux $G_3$, and in principle, no sources for $F_5$ are considered besides the flux contribution. From the Bianchi identity,
\begin{equation}
    d\star_6 de^{-4A}= -\frac{g_s}{2} G_3\wedge \bar{G}_3,
\end{equation}
where $\ast_6$ is the unwarped Hodge duality operator, $A=A(y)$, and $G_3\wedge\bar{G}_3= \frac{g_s}{6}F_3^2 \ast_6 1$. As noted in \cite{burattiDynamicaltadpolesstringy2021}, the flux contribution is not required to be a globally exact form since the internal space is not compact. From this, one concludes that
\begin{equation}
    e^{-4A}=\Lambda^2-\frac{g_s}{12}F_3^2 \, y^2,
\end{equation}
indicating that there is a relation between $F_5$ and the warping factor $A$ and $\Lambda^2$ is an integration constant. The warping factor $A$ diverges at
\begin{equation}
    y_\pm=\pm\frac{2\sqrt{3}\Lambda}{g_s^{1/2} F_3}.
\end{equation}
The distance $\Delta$ between the metric singularities $y_{\pm}$ is finite and is given by
\begin{equation}
    \Delta^2=\frac{3\Lambda^2}{g_2 F_3^2}\frac{\pi \, \Gamma^2\left(\frac{5}{4}\right)}{ \, \Gamma^2\left(\frac{7}{4}\right)}.
\end{equation}
Notice that the stronger the intensity of the flux $F_3$, the shorter the distance between singularities.\\

The non-compact space becomes compact in the presence of these two singularities. It follows that $dF_5$ is delta-like, indicating the presence of sources that must cancel the topological tadpole contribution from the fluxes; i.e., the dynamical tadpole induces the existence of 3-dimensional sources extending transversally to the former non-compact direction. This is, in fact, in agreement with the NGC conjecture, which requires that $dF_5$ be globally exact and identifies the sources as the end of the world branes (EWB) expected from the cobordism conjecture.\\

\section{\label{sec:a-double-fibration}A double fibration over the punctured sphere}
Under generic configurations, flux compactification scenarios entail a diversity of solutions to the involved equations of motion. By departing from the NGC, it becomes clear that we can fulfill the equations by demanding that they are exact global forms.\\

In this section, we shall show a non-trivial flux compactification where the final configuration and the relations among different fields that follow from the NGC are used to induce a dynamical tadpole or cobordism on a 2-dimensional plane, implying the compactification of a complex plane into a 2-dimensional sphere.\\

Let us start by reviewing a flux compactification on the compact 6-dimensional space consisting of a $\mathbb{T}^4$ fibration over a punctured sphere $\mathbb{S}^2$ corresponding to the so called model C within the framework of G-theory \cite{candelasTypeIIBflux2015, candelasTypeIIBflux2015a} (see also \cite{Damian:2016lvj}). The ten dimensional metric in the Einstein frame is given by
\begin{equation}
\label{metric}
    ds^2=e^{2A(z,\bar{z})-\phi/2}\left(\eta_{\mu\nu}dx^\mu dx^\nu +g_{ij}dy^i dy^j\right) +2e^{-2A-\phi/2}|h(z)|^2dz d\bar{z}.
\end{equation}
The torus metric, the dilaton $\phi$, and the warping factor $A$ are assumed to vary over $\mathbb{S}^2$, while the indices $i,j$ run over the $\mathbb{T}^4$ coordinates. The complex variable $z$ and $\bar{z}$ are the holomorphic and antiholomorphic coordinates over $\mathbb{S}^2$ and are given by $z=z^1+iz^2$, with $z^1, z^2\in \mathbb{R}$. The function $h(z)$ is a holomorphic function, rendering the metric well defined all over the sphere. \\

The dilaton equation in the \emph{Einstein} frame  can be written as
\begin{eqnarray}
    d\ast_6 de^{\phi}&=& e^{-\phi}de^{\phi}\wedge \ast_6 de^{\phi}-e^{-4A+\phi}de^{4A-\phi}\wedge \ast_6 de^\phi+\frac{e^{2\phi}}{2}F_3\wedge\ast F_3,
\end{eqnarray}
where we have considered $F_1=H_3=F_5=0$.
Notice that the dilaton cannot be constant, implying a dependence on internal coordinates for the flux $F_3$.\\

We can also look for a nontrivial solution in comparison with the 4-dimensional components of Einstein's equations, which read
\begin{equation}
    d\ast_6 d e^{4A-\phi}=\frac{e^{4A}}{2} F_3\wedge\ast_6 F_3.
\end{equation}

Clearly, both equations, Einstein's and dilaton's, cannot be closed forms in the presence of only a non-trivial flux $F_3$ since the flux contribution is a positively defined term. Therefore, similarly to the GKP case, we can look for conditions in which the 5-form $\ast_6 J_1$ is defined as
\begin{eqnarray}
    \ast_6 de^\phi-\ast_6 d e^{4A-\phi}+d\ast_6F_2\equiv \ast_6 J_1,
\end{eqnarray}
be trivial in cohomology. This happens for $2A=\phi$ (as found in \cite{candelasTypeIIBflux2015, candelasTypeIIBflux2015a}) with an exact and then gauged current $\ast_6J_1=d\ast_6 F_2$ implying the presence of extended objects that produce the current $\ast_6 J_1$, which can be, in principle, 3-, 5-, or 7-branes. Since we are considering a constant $C_0$ and a null $F_5$ form, the presence of $D7$ and $D3$-branes is discarded.\\


The dilaton and the Einstein equation then reduce to a single equation:
\begin{eqnarray}
    d\ast_6de^{2A}=d\ast_6 de^{\phi}&=& \frac{e^{2\phi}}{2}F_3\wedge \ast_6 F_3.
    \end{eqnarray}

Since the internal space is compact, it is clear that the presence of delta-like sources is required. Indeed, as shown in \cite{candelasTypeIIBflux2015, candelasTypeIIBflux2015a}, the RR potential $C_2$ behaves as $\log(z-z_i)$ at leading order, with $5$-dimensional extended objects (D-branes or orientifolds) located at the singular points in $S^2$, while  the holomorphic function is fixed by U-duality as
\begin{equation}
    h(z)=\frac{\eta(\sigma(z))^2\eta(\tau(z))^2}{\Pi_{i=1}^{12}(z-z_i)^{1/12}}.
\end{equation}\\
Under these circumstances, it has also been shown that this internal geometry backreacts in the presence of branes, generating a deficit angle of $\pi/6$ for each one. Hence, for the critical value of 24 branes, the complex plane curls up into a two-dimensional punctured sphere. Since we are considering only 3-form fluxes and a non-constant dilaton, only five and seven branes (the latter not carrying RR charge) will be allowed. This is precisely the framework we want to use to show that a dynamical compactification is generated by the presence of fluxes. We shall refer to these branes, although of different dimensions, as $(p,q)$-branes.\\

Notice that for smooth functions $A$ and $\phi$, the Bianchi identity is trivially satisfied, indicating that no sources are required. In such a scenario, fluxes must be independent of internal coordinates.

\subsection{Dynamical compactification}
Under the above consideration, we shall show that the complex plane compactifies into a punctured sphere in the presence of a smooth RR potential, inducing the presence of branes as expected by the cobordism conjecture. We depart from an internal non-compact space of the form $T^4\times \mathbb{C}$, where we consider only the presence of an RR field $F_3$ related to the 2-form potential
\begin{equation}
    C_2(z,\bar{z})=Nz_1 dy^1\wedge dy^4 + Mz_1 dy^2\wedge dy^3,
\end{equation}
where $z=z_1+iz_2$. The RR flux and the non-warped fiber metric are given by
\begin{eqnarray}
    F_3&=& dz^1\wedge (N dy^1\wedge dy^4+M dy^2\wedge dy^3),\\
    g&=&\text{diag } (Nz_1, Mz_1, Mz_1, Nz_1),
\end{eqnarray}
with $M,N \in \mathbb{Z}$. \\

This compactification has a peculiarity since the torus fiber becomes singular in the absence of an RR 3-form flux; in other words, the presence of a flux  supports the internal manifold in being non-singular. 

By assuming that $\phi$ and $A$ depend only on $z_2$, the Einstein and dilaton equations reduce to\footnote{We show in Appendix A, that this solution indeed corresponds to the value of the dilaton at the minimum of the corresponding potential.}
\begin{eqnarray}
    \frac{d^2\phi}{dz_2^2}=\frac{1}{z_2^2},
\end{eqnarray}
with a solution given by
\begin{equation}
    \phi=2 A=-\log(z_2)+ c_1 z_2+c_2,
\end{equation}
where $c_1$ and $c_2$ are integration constants. Hence, at leading order in small $z_2$, $\phi$ and $A$ are logarithmic, showing the presence of a singularity at least at some point in the complex plane.\\

To see explicitly how the addition of sources compactifies the complex plane into a two-sphere, let us recall the Gauss--Bonnet theorem, 
\begin{equation}
    \int_{\mathcal{M}} \kappa \, dA \;+\; \int_{\partial \mathcal{M}} \kappa_g \, ds \;=\; 2 \pi \,\chi(\mathcal{M}),
\end{equation}
where $dA$ is the area element of the surface, $ds$ is the line element along $\partial \mathcal{M}$, $\kappa$ is the Gaussian curvature, and $\kappa_g$ is the geodesic curvature of the boundary.  

Thus, for a two-dimensional surface, the Gaussian curvature is related to the Ricci scalar by 
\begin{equation}
    \kappa \;=\; \tfrac{1}{2}\,R^{(2)},
\end{equation}
where $R^{(2)}$ is the Ricci scalar associated with the two-dimensional part of the metric in the Einstein frame. In our ansatz, the component of the metric on the complex plane is
\begin{equation}
    ds^2_{(2)} \;=\; 2\,e^{-2A-\phi/2}\,|h(z)|^2\,dz\,d\bar z.
\end{equation}

This metric is more useful when written in the conformal gauge.
\begin{equation}
ds^2_{(2)} \;=\; e^{2u(z,\bar z)}\,dz\,d\bar z, 
\qquad
u(z,\bar z) \;=\; \tfrac{1}{2}\ln |h(z)|^2 \;-\; A \;-\;\tfrac{\phi}{4},
\end{equation}
which allows us to write the Gaussian curvature as
\begin{equation}
    \kappa \;=\; -\,4\,e^{-2u}\,\partial \bar\partial u,
\end{equation}
where $\partial\bar\partial = \partial_z \partial_{\bar z}$, and the factor of $4$ arises from the identity $\Delta = 4\,\partial_z \partial_{\bar z}$. Thus, the Gaussian curvature becomes negligible at points infinitely far away from the singularities. This implies that, asymptotically, the base space looks like a plane. Conversely, at the points where the base becomes singular, i.e., as $|h(z)|^2 \to \infty$, the curvature diverges. These are precisely the points that contribute to the angle deficit of the base, leading to the compactification of the complex plane. To see this effect, applying the Gauss-Bonnet theorem to the base $\mathcal{M} = \mathbb{CP}^1$ (the Riemann sphere) gives
\begin{equation}
    \int_{\mathbb{CP}^1} \kappa \, dA \;=\; 2\pi \chi(\mathbb{CP}^1)\;
\end{equation}
where the component related to the geodesic curvature at the boundary is zero, where the geodesic-curvature term vanishes because $\mathbb{CP}^{1}$ is compact and without boundary. Substituting the conformal factor,
\begin{equation}
   \int_{\mathbb{CP}^1} \kappa \, dA 
   = \,4\int_{\mathbb{CP}^1} \left[-\tfrac{1}{2}\,\partial\bar\partial\ln |h|^2
   + \partial\bar\partial(A+\tfrac{\phi}{4})\right] dz \, d\bar z.
\end{equation}
The second term integrates to zero since $A$ and $\phi$ are smooth functions, and $\partial\bar\partial$ annihilates holomorphic (and harmonic) contributions far from the singularities. However, the only nontrivial contribution arises from the divisor of the meromorphic function $h(z)$, where the Gaussian curvature diverges. In the U-duality invariant construction, this corresponds to
\begin{equation}
h(z)=\frac{\eta^{2}\!\bigl(\sigma(z)\bigr)\,
           \eta^{2}\!\bigl(\tau(z)\bigr)}
          {\displaystyle\prod_{i=1}^{n}(z-z_{i})^{1/12}}.
\label{eq:h_function}
\end{equation} 
Thus, near each zero of the denominator, the contribution of the U-duality invariant function effectively becomes a Dirac delta function, i.e., $4\,\partial \bar{\partial} \ln|z - z_i| = 2\pi\,\delta^{(2)}(z - z_i)$, so that the Gauss--Bonnet theorem reduces to
\begin{equation}
   \int_{\mathbb{CP}^1} \kappa \, dA 
   = \,\int_{\mathbb{CP}^1}\frac{\pi}{6}\sum_{i=1}^n \delta^{(2)}(z-z_i)-4\partial \bar\partial \left(\eta(\tau)^2 \eta(\sigma)^2 \right) dz d\bar z
\end{equation}
and since the Dedekind eta function is holomorphic in the upper half-plane, the second term vanishes, and the total contribution to the Gaussian curvature arises from the singularities of the complex plane where the branes are located. Consequently, we obtain
\begin{equation}
   \int_{\mathbb{CP}^1} \kappa \, dA \;=\;
   2\pi \sum_{i=1}^n \frac{1}{12}.
\end{equation}

Thus, it is concluded that each puncture contributes an angle deficit of $\delta=\pi/6$, i.e. $1/12$ of the Euler characteristic. And precisely for $n=24$ of such singularities, the total deficit is $4\pi$, reproducing the Euler characteristic $\chi(\mathbb{CP}^1)=2$. 

\section{From (co)homology to cobordism}
\label{sec:cohomology-to-cobordism}
The requirement to have 24 singularities in two blocks of 12 each is a direct consequence of geometry and dualities at the level of the equations of motion, as shown above and generically studied in \cite{candelasTypeIIBflux2015, candelasTypeIIBflux2015a}. Therefore, it is expected that such an issue arises as a result of computing the corresponding (co)homology group. In this section, we present this calculation as well as the extension to compute the spin cobordism group related to our scenario in order to determine, according to the cobordism conjecture, whether more structure and extended objects are required to cancel the group.\\


In the Model C configuration of G-theory, the effective four-dimensional vacuum arises from Type IIB string theory compactified on a manifold locally diffeomorphic to $T^4 \times \mathbb{C}$ with non-trivial RR 3-form flux $F_3$. As established in Section \ref{sec:gkp-warped}, the U-duality group acting on the moduli space for the $n=2$ truncation is the product of modular groups $G = SL(2,\mathbb{Z})_\tau \times SL(2,\mathbb{Z})_\sigma$. To determine the topological obstructions to this vacuum, we must calculate the spin bordism group of the classifying space $BG$, specifically $\Omega_6^{\text{Spin}}(BG)$\footnote{Throughout we compute the bordism of the bosonic, ungauged duality group $G = SL(2,\mathbb{Z})_\tau \times SL(2,\mathbb{Z})_\sigma$ with an ordinary Spin structure; this is the frame in which all results of Sections~\ref{sec:bordism}--\ref{sec:homvsbord} are stated. The exchange symmetry $\tau \leftrightarrow \sigma$ enters the main text only as an \emph{automorphism} of this ungauged computation, it permutes the $(1,5)$ and $(5,1)$ contributions and organizes the generators of~Eq.\eqref{eq:generators},which requires no gauging and is independent of the two refinements below. First, if $\tau \leftrightarrow \sigma$ is gauged, $G$ is enlarged to the wreath product $G' = SL(2,\mathbb{Z}) \wr \mathbb{Z}_2 = (SL(2,\mathbb{Z}) \times SL(2,\mathbb{Z})) \rtimes \mathbb{Z}_2$. At odd primes the transfer identifies the result with the $\mathbb{Z}_2$-invariants of $\Omega^{Spin}_6(BG)$: the $(1,5)$ and $(5,1)$ classes combine into a single $\mathbb{Z}_3$, while the $(3,3)$ class lies in the sign representation, the swap of two degree-$3$ factors carries a Koszul sign $(-1)^{3\cdot 3}=-1$, and is projected out; the $2$-primary part is a separate, harder problem not given by invariants. As a physical consequence, the mixed $(3,3)$ charge identified in Section~\ref{sec:homvsbord} as a genuine $G$-theory effect survives precisely because $\tau \leftrightarrow \sigma$ is treated as a global symmetry; it would be absent were the exchange gauged. Second, including fermions replaces each $SL(2,\mathbb{Z})$ by its metaplectic cover; since $Mp(2,\mathbb{Z})$ is a $\mathbb{Z}_2$ central extension of $SL(2,\mathbb{Z})$ it is invisible at odd primes, so the odd-primary torsion, in particular $(\mathbb{Z}_3)^3$, is unchanged and only the $2$-primary sector is modified. The two covers share a single fermion parity $(-1)^F$, refining $G$ to the central product $\widetilde{G} = Mp(2,\mathbb{Z}) \times_{\mathbb{Z}_2} Mp(2,\mathbb{Z})$, with tangential structure $(Spin \times \widetilde{G})/\mathbb{Z}_2$ and twisted bordism $\Omega^{Spin\text{-}\widetilde{G}}_6$. Both refinements are well posed and left for future work.}. All details about this computation are shown in Appendix \ref{appendixB}.\\

\subsection{Homology of the Duality Group}

The classifying space $BG$ decomposes as the product $BSL(2,\mathbb{Z}) \times BSL(2,\mathbb{Z})$. We first recall the integral homology of the modular group $SL(2,\mathbb{Z}) \cong \mathbb{Z}_4 *_{\mathbb{Z}_2} \mathbb{Z}_6$. The homology groups $H_k(BSL(2,\mathbb{Z}); \mathbb{Z})$ are well-known torsion groups for $k > 0$:
\begin{equation}\label{eq:homologiaSL2Z}
    H_k(BSL(2,\mathbb{Z}); \mathbb{Z}) \cong 
    \begin{cases} 
    \mathbb{Z} & k=0 \\
    \mathbb{Z}_{12} & k \text{ odd} \\
    0 & k \text{ even, } k>0 
    \end{cases}
\end{equation}
To determine the homology of the full duality group $BG$, we apply the Künneth formula. Retaining the $\text{Tor}(\mathbb{Z}_{12},\mathbb{Z}_{12})\cong\mathbb{Z}_{12}$ terms that arise whenever both indices are odd, the relevant terms are given by

\begin{eqnarray}\label{eq:homologiaBG}
H_0(BG)&\cong&\mathbb{Z},\nonumber\\
H_1(BG)&\cong&(\mathbb{Z}_{12})^2,\nonumber\\
H_2(BG)&\cong& \mathbb{Z}_{12},\nonumber\\
H_3(BG)&\cong& (\mathbb{Z}_{12})^3,\nonumber\\
H_4(BG)&\cong& (\mathbb{Z}_{12})^2,\nonumber\\
H_5(BG)&\cong& (\mathbb{Z}_{12})^4,\nonumber\\
H_6(BG)&\cong& (\mathbb{Z}_{12})^3.
\end{eqnarray}

Before going on, it is important to note that the homology group
reproduces what we have encountered at the level of the equations of motion, as expected. In this sense, the cancellation of the (co)homological global charge by adding two sets of 12 branes leads us to consider only the zero class in (co)homology. The next step is to compute the bordism group and see under which conditions the group vanishes according to the cobordism conjecture. We closely follow \cite{Debray:2023yrs}.\\

\subsection{\label{sec:bordism}Bordism group}

After using the Atiyah-Hirzebruch sequence as shown in Appendix \ref{appendixB}, we find that
\begin{equation} 
\Omega_6^{\mathrm{Spin}}\bigl(B(SL(2,\mathbb Z)\times SL(2,\mathbb Z))\bigr)
\cong
\Omega_6^{\mathrm{Spin}}\bigl(B(SL(2,\mathbb Z)\times SL(2,\mathbb Z))\bigr)_{(2)}
\oplus
(\mathbb{Z}_3)^3,
\end{equation}
where, using the $ko$-homology input of Ref.~\cite{Barcenas},\footnote{Theorem~5.8 of the first version of Ref.~\cite{Barcenas} states $\log_2|\widetilde{ko}_6(B\mathbb{Z}_4\wedge B\mathbb{Z}_4)|=4$, whereas the $E_\infty$-page displayed in their Figure~5.3 has five $\mathbb{Z}_2$ summands in total degree $6$, giving $\log_2|\widetilde{ko}_6(B\mathbb{Z}_4\wedge B\mathbb{Z}_4)|=5$. We thank the authors for confirming that the figure is correct and that the theorem contained a typo, which will be corrected in a forthcoming version.}
$\Omega_6^{\mathrm{Spin}}\bigl(B(SL(2,\mathbb Z)\times SL(2,\mathbb Z))\bigr)_{(2)}$
has order $32$. Imposing that the $2$-primary part has exponent dividing $4$ leaves three a priori candidates:
\begin{equation} \label{eq:candidates}
    \mathbb{Z}_4\oplus\mathbb{Z}_4\oplus\mathbb{Z}_2,\quad
    \mathbb{Z}_4\oplus\mathbb{Z}_2\oplus\mathbb{Z}_2\oplus\mathbb{Z}_2,\quad
    (\mathbb{Z}_2)^5.
\end{equation}
We now argue, in three steps, that (i)~the $2$-primary part has an exponent dividing $4$, (ii)~$(\mathbb{Z}_2)^5$ is excluded, and (iii)~explicit generating manifolds select $\mathbb{Z}_4 \oplus (\mathbb{Z}_2)^3$ among the three candidates given in Eq.~\eqref{eq:candidates}. This reinforces the idea that physics at high energies requires a more elaborate mathematical structure that contains the structures related to low energies: here, the bordism group strictly contains the cohomological information constructible from the equations of motion at the
perturbative level.\\

We now bound the exponent of the $2$-primary part. As established in Appendix~\ref{appendixB}, the inclusion $B\mathbb{Z}_4 \to B\,SL(2,\mathbb{Z})$ is a
$2$-local homology equivalence, so at the prime $2$ the relevant groups are computed from the smash product $\mathrm{ko}f_6(B\mathbb{Z}_4 \wedge B\mathbb{Z}_4)$, which
captures the interaction between the two $SL(2,\mathbb{Z})$ factors. The reduced $\mathrm{ko}$-homology of $B\mathbb{Z}_4$ is not annihilated by $4$ in general the explicit computation of Ref.~\cite{Barcenas} gives factors such as $\widetilde{\mathrm{ko}}_{8d+3}(B\mathbb{Z}_4)=\mathbb{Z}_{2^{4d+3}}\oplus\mathbb{Z}_{2^{2d+1}}$ but in the low degrees feeding total degree $6$ the factors have exponent dividing $4$: $\widetilde{\mathrm{ko}}_1(B\mathbb{Z}_4)=(\mathbb{Z}_2)^2$, $\widetilde{\mathrm{ko}}_2=\mathbb{Z}_2$, $\widetilde{\mathrm{ko}}_3$ of order $4$, and $\widetilde{\mathrm{ko}}_5=\mathbb{Z}_2$. These factors assemble into
$\mathrm{ko}f_6(B\mathbb{Z}_4 \wedge B\mathbb{Z}_4)$ through the $\mathrm{ko}$-module structure of the smash, and Ref.~\cite{Barcenas} determines that $|\mathrm{ko}f_6(B\mathbb{Z}_4 \wedge B\mathbb{Z}_4)|=2^5=32$. The full isomorphism type of this group is not given there; we take as a working hypothesis, consistent with the
generators of Section~\ref{sec:bordism} below, that no hidden $h_0$-extension raises the exponent, so the $2$-primary part has exponent dividing $4$. Under this hypothesis, the abelian groups of order $32$ and exponent dividing $4$ are exactly the three listed in Eq.~\eqref{eq:candidates}.\\

We record the role of the exchange symmetry separately. The
$\tau \leftrightarrow \sigma$ symmetry permutes the two factors and acts on $\mathrm{ko}f_6(B\mathbb{Z}_4 \wedge B\mathbb{Z}_4)$; at degree $6$ the three contributing pairs of the Landweber formula, $(1,5)$, $(3,3)$ and $(5,1)$, are
organized by it, with $(1,5)$ and $(5,1)$ exchanged and $(3,3)$ fixed. This constrains how the candidate summands are paired and will be used below to organize the generators. It is, however, the annihilation-by-$4$ property above (not the symmetry) that bounds the exponent; the symmetry alone would
only pair summands, not forbid an order-$8$ element.

The candidate $(\mathbb{Z}_2)^5$ has exponent $2$ and is therefore excluded as soon as a single element of order $4$ is exhibited. One might expect such an element already at the $E_2$ page, in the entry $E^2_{2,4}=H_2(BG;\Omega^{\mathrm{Spin}}_4(\mathrm{pt}))\cong \mathbb{Z}_{12}$, whose $2$-primary part is $\mathbb{Z}_4$ and which encodes the coupling of the gravitational Pontryagin class (arising from $\Omega^{\mathrm{Spin}}_4$) to the U-duality flux. It is, however, $E^\infty_{2,4}$ rather than $E^2_{2,4}$ that is a subquotient of the final group, and the generator analysis below reduces this entry to $\mathbb{Z}_2$: it is generated by $K3\times L^1_4\times L^1_4$, of order $2$. The order-$4$ element, therefore, originates from the pure-monodromy class $L^3_4\times L^3_4$ of \eqref{eq:generators}, so that the exclusion of $(\mathbb{Z}_2)^5$ rests on the order of $L^3_4\times L^3_4$, the same input that distinguishes the two remaining candidates.
\begin{equation}
\mathbb{Z}_4\oplus\mathbb{Z}_2\oplus\mathbb{Z}_2\oplus\mathbb{Z}_2
\quad\text{and}\quad
\mathbb{Z}_4\oplus\mathbb{Z}_4\oplus\mathbb{Z}_2,
\label{eq:two-candidates}
\end{equation}
which differ only in the number of independent $\mathbb{Z}_4$ summands.\\

The last ambiguity is resolved by exhibiting explicit generating manifolds. Let $L^n_k = S^n/(\mathbb{Z}_k)$ denote the lens space and let $Q^n_k$ denote the lens-space bundle over $\mathbb{CP}^1$ of Ref.~\cite{Debray:2023yrs}; the manifolds $L^1_4$, $L^3_4$ and $Q^5_4$ are Spin and carry canonical $\mathbb{Z}_4$ bundles from the quotients defining them. The $2$-primary part is generated by
\begin{equation}
L^3_4 \times L^3_4,\quad
Q^5_4 \times L^1_4,\quad
L^1_4 \times Q^5_4,\quad
K3 \times L^1_4 \times L^1_4,
\label{eq:generators}
\end{equation}
each equipped with the product $\mathbb{Z}_4 \times \mathbb{Z}_4$ bundle (trivial along the $K3$ factor). The generator analysis indicates that the class $L^{3}_{4}\times L^{3}_{4}$ has order four, generating the single $\mathbb{Z}_{4}$ summand, while the remaining three classes have order two; the two classes $Q^{5}_{4}\times L^{1}_{4}$ and
$L^{1}_{4}\times Q^{5}_{4}$ are distinct because they distinguish the two duality factors. Under these order assignments no second order-four class appears, and the surviving group is $\mathbb{Z}_{4}\oplus(\mathbb{Z}_{2})^{3}$ rather than $\mathbb{Z}_{4}\oplus\mathbb{Z}_{4}\oplus\mathbb{Z}_{2}$. We stress that these
assignments, in particular the absence of the hidden extension $2\,[L^{3}_{4}\times L^{3}_{4}]=[K3\times L^{1}_{4}\times L^{1}_{4}]$, remain to be established by the secondary ($\rho$-type) bordism invariant of footnote~\ref{fn:rho}, which we do not compute here. All statements of the form $\Omega^{\mathrm{Spin}}_{6}(BG)_{(2)}\cong\mathbb{Z}_{4}\oplus(\mathbb{Z}_{2})^{3}$ in what follows are to be read subject to this caveat. Note that the surviving
$\mathbb{Z}_4$ is generated by the pure-monodromy class $L^3_4 \times L^3_4$, whereas the gravitational class $K3 \times L^1_4 \times L^1_4$ sitting at
$E^2_{2,4}$ contributes only a $\mathbb{Z}_2$. The generators are moreover compatible with the exchange symmetry anticipated above: it permutes $Q^5_4 \times L^1_4 \leftrightarrow L^1_4 \times Q^5_4$, acts on $L^3_4 \times L^3_4$ by inversion of the $\mathbb{Z}_4$ generator (with the Koszul sign of interchanging two odd-dimensional factors), and fixes $K3 \times L^1_4 \times L^1_4$.\\

To summarize, the full bordism group is
\begin{equation} 
\Omega^{Spin}_6\big(B(SL(2,\mathbb{Z}) \times SL(2,\mathbb{Z}))\big)
\;\cong\;
\big(\mathbb{Z}_4 \oplus (\mathbb{Z}_2)^3\big) \oplus (\mathbb{Z}_3)^3,
\label{eq:cobordism_group}
\end{equation}

where the $2$-primary part is the one indicated by the generators of
Eq.~\eqref{eq:generators}, pending the $\eta$/$\rho$-invariant computation
described above.\footnote{\label{fn:rho} The two surviving candidates differ by whether there is a hidden extension $2\,[L^3_4\times L^3_4]=[K3\times L^1_4\times L^1_4]$ in the Atiyah--Hirzebruch filtration. The generator analysis indicates that $[K3\times L^1_4\times L^1_4]$ has order $2$ independently, giving \eqref{eq:cobordism_group}. A rigorous proof requires a secondary ($\rho$-type) invariant, which we do not compute here, valued in $\mathbb{Q}/\mathbb{Z}$: for a Spin coboundary $W^{7}$ with $\partial W^{7}=k\,(L^3_4\times L^3_4)$ one sets $\rho=\tfrac1k\big(\int_{W^{7}}\hat A(W)\,\mathrm{ch}(L_1\otimes L_2)-\mathrm{ind}_{\rm APS}\big)$, whose value on the two order-two classes decides the extension.}\\

\subsection{\label{sec:homvsbord}Homology versus Bordism: the need for non-perturbative objects}

One of the central results of this paper is that the bordism group give in Eq.~\eqref{eq:cobordism_group}
contains strictly more structure than the (co)homological global charge computed from
the equations of motion. We now make this contrast precise and spell out its physical
consequences.\\

We have seen that $H_2(BG;\mathbb{Z})\;\cong\;\mathbb{Z}_{12}$.
At the prime $3$, this yields a single $\mathbb{Z}_3$; at the prime $2$ it yields a single
$\mathbb{Z}_4$. Since $\mathbb{Z}_{12}\cong \mathbb{Z}_3\oplus\mathbb{Z}_4$, physically, this is the charge measured by the equations of motion: the
24 $(p,q)$-branes, organized in two blocks of $12$, precisely cancel this cohomological
tadpole. This is the perturbative consistency condition: the content of the GKP tadpole
cancelation reinterpreted through the lens of generalized symmetries.\\

On the other hand, the bordism group given in Eq.~\eqref{eq:cobordism_group} is markedly larger.
At the prime $3$, homology gives one $\mathbb{Z}_3$,
          whereas bordism gives $(\mathbb{Z}_3)^3$.
    At the prime $2$ homology gives a group of order $4$, whereas
          bordism gives a group of order $\mathbf{32}$.\\

The origin of the two extra $\mathbb{Z}_3$ factors is transparent from the 
K\"unneth formula~\eqref{eq:Kunneth}. The three pairs $(i,j)$ with $i+j=6$ contributing
at the prime $3$ are $(1,5)$, $(3,3)$, and $(5,1)$. The pairs $(1,5)$ and $(5,1)$ arise
from each $\mathrm{SL}(2,\mathbb{Z})$ factor independently and have direct analogs in the
F-theory cobordism. The pair $(3,3)$, however, is a mixed contribution that
requires both $\mathrm{SL}(2,\mathbb{Z})$ factors to be simultaneously non-trivial.
This mixed $\mathbb{Z}_3$ charge has no F-theory analog. It is a genuinely
G-theory effect, encoding a topological obstruction that couples to both moduli $\tau$
and $\sigma$ simultaneously.\\

\section{\label{sec:conclusions}Conclusions and final comments}
The cobordism conjecture states that the cobordism group of any consistent quantum gravity theory must be trivial. In our setup, the $(p,q)$-branes visible at the perturbative level cancel the cohomological charge $H_2(BG;\mathbb{Z})\cong\mathbb{Z}_{12}$. However, the bordism group given in Eq.~\eqref{eq:cobordism_group} is strictly finer than the cohomological charge: the class $H_{2}(BG;\mathbb{Z})\cong\mathbb{Z}_{12}$ enters $\Omega^{\mathrm{Spin}}_{6}(BG)$ only through the subquotient $E^{\infty}_{2,4}$ of the Atiyah--Hirzebruch filtration, and
\begin{equation}
\left|\Omega^{\mathrm{Spin}}_{6}(BG)\right| \;=\; 32\cdot 27
\;\;\gg\;\;
\left|H_{2}(BG;\mathbb{Z})\right| \;=\; 12 .
\label{eq:refinement}
\end{equation}
Indeed, the generator analysis of Section~\ref{sec:bordism} indicates that the $2$-primary part of the homological charge survives only as the order-two class
$[K3\times L^{1}_{4}\times L^{1}_{4}]$, so the refinement is not a subgroup inclusion: the residual torsion charges of \eqref{eq:cobordism_group} are not canceled by the $24$ $(p,q)$-branes alone. The consistency of the G-theory vacuum therefore requires the existence of additional non-perturbative objects that carry and absorb the extra torsion charges. We identify these as follows:

\begin{itemize}
    \item \textit{Extra $(\mathbb{Z}_3)^2$ charges}: These are associated with the order-three elements of the duality group. The
combination $ST$ of S- and T-duality satisfies $(ST)^{3}=-\mathbb{1}$ and hence
has order six in $\mathrm{SL}(2,\mathbb{Z})$, projecting to an order-three
element of $\mathrm{PSL}(2,\mathbb{Z})$; the order-three elements of
$\mathrm{SL}(2,\mathbb{Z})$ itself are generated by $(ST)^{2}$, consistent with
the amalgam structure
$\mathrm{SL}(2,\mathbb{Z})\cong\mathbb{Z}_{4}*_{\mathbb{Z}_{2}}\mathbb{Z}_{6}$.
    The bordism analysis suggests the presence of non-geometric defects implementing this $\mathbb{Z}_3$
    monodromy, known in the literature as $\mathbb{Z}_3$ S-folds~\cite{Aharony:2016kai}.
    These are invisible in string perturbation theory, and our analysis suggests they cannot be detected by the
    equations of motion or by the Gauss-Bonnet analysis.

   \item \textit{Extra 2-primary charges}: At the prime $2$ the cohomological tadpole
    sees only the $\mathbb{Z}_4\subset H_2(BG;\mathbb{Z})$, whereas the bordism group
    contributes $\mathbb{Z}_4\oplus(\mathbb{Z}_2)^3$ of order $32$. The structure beyond
    the homological $\mathbb{Z}_4$ therefore has order $8$ and is carried by the three
    order-two classes $[Q^5_4\times L^1_4]$, $[L^1_4\times Q^5_4]$ and
    $[K3\times L^1_4\times L^1_4]$ of~\eqref{eq:generators}. The first two require both
    $\mathrm{SL}(2,\mathbb{Z})$ factors to act non-trivially and have no single-factor
    analog in F-theory; we interpret them as defects coupling simultaneously to the
    S-duality monodromies of the two tori, the 2-primary counterpart of the mixed
    $(3,3)$ charge identified above. The class $[K3\times L^1_4\times L^1_4]$ couples the
    gravitational Pontryagin class to a single duality flux  and is the bordism image of the $E^2_{2,4}$ entry; it is expected to carry order two, since $\hat{A}(K3)=2$ contributes an even factor to its detecting invariant. The surviving $\mathbb{Z}_4$ is instead generated by the pure-monodromy class $[L^3_4 \times L^3_4]$, whose order the same $\eta$-invariant computation would fix at four.
\end{itemize}

This observation constitutes one of the principal physical results of our work. The
structure of the bordism group reveals that the G-theory vacuum is not fully
consistent at the level of tadpole cancellation alone. The cobordism conjecture predicts
the existence of non-perturbative defects, S-folds, and mixed U-duality defects that
are invisible to the perturbative equations of motion but are required by the global
topological consistency of the vacuum. In this sense, cobordism is a strictly stronger
consistency condition than tadpole cancellation in the G-theory setting.\\

We stress that this conclusion depends crucially on the product structure
$G=\mathrm{SL}(2,\mathbb{Z})\times\mathrm{SL}(2,\mathbb{Z})$ of the U-duality group.
For a single $\mathrm{SL}(2,\mathbb{Z})$ factor, as in F-theory, the analogous
bordism group is smaller and closer to the homological result. The extra structure
in~\eqref{eq:cobordism_group} is therefore a direct consequence of the richer duality
group of G-theory and provides a topological fingerprint distinguishing G-theory vacua
from their F-theory counterparts.\\

Manifestly, these results reinforce the theoretical structure expected from a consistent physical theory. As a consequence, the model successfully incorporates lower-energy laws within a perturbative framework. Regarding the equations of motion, cohomology proves sufficient to reproduce those derived from the action. This correspondence, however, reaches its limit when exploring the richer structures at higher energies. Hence, such complexity demands a transition toward more sophisticated mathematical foundations. And it is precisely here that (co)bordisms provide the necessary framework to extend our understanding.

\begin{center}
    {\bf Acknowledgments}
\end{center}
 C.D. is supported by SECIHTI Frontier Science CF-2023-I-682 while O. L-B is supported by CIIC-UG-DAIP No. 236/2022 and CBF-2025-I-2708. V.M,L.-R. is supported by a SECIHTI doctoral fellowship. We are grateful to Arun Debray for illuminating discussions on our results and to L. E. García-Hernández for his comments concerning the group dimensions.

\appendix
\section{Fixing the dilaton through scalar potential}
Following the metric ansatz introduced in Eq.~\eqref{metric}, the scalar potential in the four-dimensional effective theory receives contributions from the fluxes, the dilaton, and the internal curvature. 
The background geometry is locally of the form $T^4\times C$, with complex coordinate $z = x + i z_2$ on the base and holomorphic functions parametrizing the fluxes and warping as in the Type IIB constructions of~\cite{candelasTypeIIBflux2015,burattiDynamicalCobordismSwampland2021,blumenhagenDynamicalcobordismdomain2022}. 

In the Type IIB theory, the contribution from the three-form flux $G_3=F_3 - \tau H_3$ to the scalar potential is
\begin{equation}
    V_{\text{flux}} = \frac{1}{4\kappa_{10}^2}\int \frac{G_3 \wedge *\bar{G}_3}{\text{Im}\,\tau}\,.
\end{equation}
For the present ansatz, we consider a purely RR flux background ($H_3=0$), such that $G_3=F_3$. 
The flux contribution then simplifies to
\begin{equation}
    V_{\text{flux}} = \frac{1}{2\kappa_{10}^2}\int \frac{MN}{2s}\, d\text{vol}_4\,,
\end{equation}
where $s=e^{-\phi}$ and $d\text{vol}_4=\sqrt{-g_4}\,dx^0\wedge dx^1\wedge dx^2\wedge dx^3$. 
This term corresponds to the flux-induced potential familiar from flux compactifications, where $(M,N)$ are the effective flux quanta threading the internal torus.

The kinetic term for the dilaton field also generates a contribution to the scalar potential, given in the ten-dimensional Einstein frame by
\begin{equation}
    V_{\text{dilaton}} = \frac{1}{4\kappa_{10}^2}\int d\phi\wedge *d\phi\,,
\end{equation}
where $*$ denotes the Hodge dual in ten dimensions.
Assuming the dilaton depends only on the complex coordinate $z$ and its conjugate $\bar{z}$,
\begin{equation}
    d\phi = \partial\phi\,dz + \bar{\partial}\phi\,d\bar{z}\,,
\end{equation}
its dual is
\begin{equation}
    *d\phi = \frac{e^{2A}}{|h(z)|^2}\big(\partial\phi\,d\bar{z} - \bar{\partial}\phi\,dz\big)\wedge d\text{vol}_8\,,
\end{equation}
where $e^{2A}$ is the warp factor and $|h(z)|$ is the local holomorphic density introduced in~\cite{candelasTypeIIBflux2015}.  
This yields the contribution
\begin{equation}
    V_{\text{dilaton}} = \frac{1}{2\kappa_{10}^2}\int \frac{(s')^2}{2s}\,d\text{vol}_4\,,
\end{equation}
where $s' = \partial_{z_2}s$ denotes the variation of the inverse string coupling along the transverse direction. The gravitational term is more intricate, as the ansatz includes a non-trivial warping factor.
The Ricci scalar can be decomposed as
\begin{equation}
    R = R^{(4)} + R^{(6)}\,.
\end{equation}
For the internal curvature $R^{(6)}$, and assuming the warp factor and complex moduli satisfy $e^{-2A}=(\tau_2\sigma_2)^{1/2}$ with $\tau_2=M z_2$ and $\sigma_2=N z_2$, we obtain
\begin{equation}
    R^{(6)} = \frac{e^{\phi/2}}{8}\frac{1}{(MN)^{1/2}}\frac{1}{z_2^3 |h|^2}\left(4 + 6z_2\phi' + 5z_2^2(\phi')^2 - 7z_2^2\phi''\right),
\end{equation}
where $\phi' = d\phi/dz_2$. 
Using $e^{-\phi}=s$, one finds $\phi'=-s'/s$ and $\phi''=(s')^2/s^2 - s''/s$. 
Substituting these into the above expression gives the gravitational contribution to the potential,
\begin{equation}
    V_{\text{grav}} = \frac{1}{2\kappa_{10}^2}\int \frac{1}{4}\left[\frac{4s}{z_2^2} - 6\frac{s'}{z_2} - 2\frac{(s')^2}{s} + 7s''\right] d\text{vol}_4\,.
\end{equation}
The extremization of the total scalar potential with respect to $s$ then yields
\begin{equation}
    s = \sqrt{MN}\,z_2\,,
\end{equation}
which corresponds to the expected on-shell dilaton profile. 
At this value, the dilaton equation of motion is satisfied, and the potential reaches its minimum, consistent with the dynamical resolution of the tadpole proposed in~\cite{blumenhagenDynamicalcobordismdomain2022}.\\

In the derivation above, the variations $s'$ and $s''$ are treated as independent off-shell quantities.
This approach is standard in the dimensional reduction of higher-derivative scalar-tensor actions, where the potential is computed before imposing the field equations. 
Physically, $s'$ encodes the spatial variation of the string coupling along the internal coordinate $z_2$, while $s''$ characterizes its curvature in field space.
Treating them independently allows the identification of extremal configurations that satisfy the ten-dimensional equations of motion upon variation.\\

This method is consistent with the framework of local dynamical cobordism~\cite{blumenhagenDynamicalcobordismdomain2022}, where spacetime-dependent scalar fields evolve along a finite interval capped by an end-of-the-world brane.
At the extremum of the effective potential, the field gradients adjust such that the full configuration satisfies both the dilaton and Einstein equations. 
Thus, although $s'$ and $s''$ are dynamically linked on-shell, considering them as independent fields off-shell is a justified and necessary step in deriving the consistent scalar potential governing these dynamical flux backgrounds.\\

\section{Computation of $\Omega_6^{\mathrm{Spin}}(BG)$}
\label{appendixB}

In this section, we describe the computation of the 6th Spin bordism group of the space
\[
BG=B\left(SL(2,\mathbb{Z})\times SL(2,\mathbb{Z})\right).
\]
The main tool will be the Atiyah-Hirzebruch spectral sequence. We first construct the $E^2$ page in detail, and then we use a decomposition into primary parts, following a strategy similar to the one used in \cite{Debray:2023yrs}.\\

The Atiyah-Hirzebruch spectral sequence for Spin bordism converges to $\Omega^{\mathrm{Spin}}_{p+q}(BG)$ and has an initial page
\begin{equation}
E^{2}_{p,q}=H_p\bigl(BG;\Omega_q^{\mathrm{Spin}}(\mathrm{pt})\bigr),
\qquad p,q\geq 0.
\label{eq:AHSS_E2}
\end{equation}
Its differentials are of the form
\begin{equation}
d_r:E^r_{p,q}\longrightarrow E^r_{p-r,q+r-1}.
\label{eq:AHSS_dr}
\end{equation}

Once the spectral sequence stabilizes, we obtain the $E^\infty$ page, which induces a filtration of $\Omega^{\mathrm{Spin}}_6(BG)$. In particular, the group we are looking for is reconstructed from the diagonal
\[
p+q=6
\]
on the final page. To construct the $E^2$ page, we need to compute the groups
\[
H_p\bigl(BG;\Omega_q^{\mathrm{Spin}}(\mathrm{pt})\bigr).
\]
The procedure will be as follows:
\begin{enumerate}
\item First, we compute the integral homology $H_p(BG;\mathbb{Z})$ from the homology of $BSL(2,\mathbb{Z})$, using the Künneth formula.
\item Then, for each $q$, there is a mapping
\[H_p(BG;\mathbb{Z})\rightarrow H_p(BG;\Omega_q^{\mathrm{Spin}}(\mathrm{pt})),\]
which follows from the universal coefficient theorem.
\end{enumerate}

From the physical point of view, when we study a theory with a discrete symmetry or duality group $G$, the corresponding topological backgrounds are described by principal $G$-bundles over spacetime. These bundles capture the global topological information of the background, beyond its local description. A standard fact in algebraic topology is that principal $G$-bundles over a space $X$ are classified, up to isomorphism, by homotopy classes of maps
\[
X\to BG,
\]
where $BG$ is the classifying space of $G$. Equivalently, every principal $G$-bundle can be obtained as the pullback of the universal bundle over $BG$. This is why the relevant bordism groups naturally take the form
\[
\Omega_*^{\mathrm{Spin}}(BG).
\]
Indeed, an element of $\Omega_n^{\mathrm{Spin}}(BG)$ is represented by an $n$-dimensional Spin manifold together with a map to $BG$, and this map encodes precisely the topological $G$-background carried by the manifold.

In our case,
\[
G=SL(2,\mathbb{Z})\times SL(2,\mathbb{Z}),
\]
so the space that enters the spectral sequence is
\[
BG=B\left(SL(2,\mathbb{Z})\times SL(2,\mathbb{Z})\right).
\]
Therefore, the groups we want to study are the Spin bordism groups of manifolds equipped with a topological background for this duality group.

\subsection{Computation of $H_p(BG;\mathbb{Z})$}

Let us denote $X=B(SL(2,\mathbb{Z}))$.
Using the homology groups of $BSL(2,\mathbb{Z})$ in
Eq.~\eqref{eq:homologiaSL2Z} together with the Künneth formula
\begin{equation}
0\longrightarrow
\bigoplus_{p+q=n}
\left(
H_p(X;\mathbb{Z})\otimes H_q(X;\mathbb{Z})
\right)
\longrightarrow
H_n(X\times X;\mathbb{Z})
\longrightarrow
\bigoplus_{p+q=n-1}
\operatorname{Tor}^{\mathbb{Z}}_1
\left(
H_p(X;\mathbb{Z}),H_q(X;\mathbb{Z})
\right)
\longrightarrow 0,
\label{eq:Kunneth}
\end{equation}
the low-degree homology groups of $BG$ listed in
Eq.~\eqref{eq:homologiaBG} are obtained.\\

The tensor part $\bigoplus_{p+q=n}H_p(X)\otimes H_q(X)$ in \eqref{eq:Kunneth} contributes a $\mathbb{Z}_{12}$ for each pair with one index zero or both indices odd, while the torsion part $\bigoplus_{p+q=n-1}\mathrm{Tor}(H_p(X),H_q(X))$ adds a further $\mathbb{Z}_{12}$ for each pair of odd indices summing to $n-1$. Since a sum of two odd numbers is even, the latter is nonzero only for odd $n$: it contributes one extra $\mathbb{Z}_{12}$ to $H_3(BG)$, from the pair $(1,1)$, two to $H_5(BG)$, from $(1,3)$ and $(3,1)$, and nothing in even degree.\\

The next step is to use these groups to fill the non-trivial rows of the $E^2$ page. Recall that, in low degrees, the Spin bordism groups of the point are given by
\begin{equation}
\Omega_q^{\mathrm{Spin}}(\mathrm{pt})\cong
\begin{cases}
\mathbb{Z}, & q=0,4,\\
\mathbb{Z}_2, & q=1,2,\\
0, & q=3,5,6.
\end{cases}
\label{eq:Spinpt}
\end{equation}
Thus, for each fixed value of $q$, the entries of the $E^2$ page are obtained by computing
\[
H_p\bigl(BG;\Omega_q^{\mathrm{Spin}}(\mathrm{pt})\bigr).
\]
For this purpose, we use the universal coefficient theorem
\begin{equation}
0\to H_p(BG;\mathbb{Z})\otimes \Omega_q^{\mathrm{Spin}}(\mathrm{pt})\to
H_p(BG;\Omega_q^{\mathrm{Spin}}(\mathrm{pt}))\to
\mathrm{Tor}\bigl(H_{p-1}(BG;\mathbb{Z}),\Omega_q^{\mathrm{Spin}}(\mathrm{pt})\bigr)\to 0.
\label{eq:UCT}
\end{equation}

From this, we conclude that
\begin{equation}
H_p\bigl(BG;\Omega_0^{\mathrm{Spin}}(\mathrm{pt})\bigr)\cong H_p\bigl(BG;\Omega_4^{\mathrm{Spin}}(\mathrm{pt})\bigr)\cong H_p(BG;\mathbb{Z}),
\end{equation}
since
\[
\Omega_0^{\mathrm{Spin}}(\mathrm{pt})\cong
\Omega_4^{\mathrm{Spin}}(\mathrm{pt})\cong \mathbb{Z},
\]
and $\mathrm{Tor}(-,\mathbb{Z})=0$.

Similarly, since
\[
\Omega_1^{\mathrm{Spin}}(\mathrm{pt})\cong
\Omega_2^{\mathrm{Spin}}(\mathrm{pt})\cong \mathbb{Z}_2,
\]
the previous short exact sequence allows us to compute the entries with coefficients in $\mathbb{Z}_2$, from which we obtain
\begin{itemize}
    \item $H_0(BG;\mathbb{Z}_2)\cong \mathbb{Z}_2,$
    \item $H_1(BG;\mathbb{Z}_2)\cong (\mathbb{Z}_2)^2,$
    \item $H_2(BG;\mathbb{Z}_2)\cong (\mathbb{Z}_2)^3,$
    \item $H_3(BG;\mathbb{Z}_2)\cong (\mathbb{Z}_2)^4,$
    \item $H_4(BG;\mathbb{Z}_2)\cong (\mathbb{Z}_2)^5,$
    \item $H_5(BG;\mathbb{Z}_2)\cong (\mathbb{Z}_2)^6,$
    \item $H_6(BG;\mathbb{Z}_2)\cong (\mathbb{Z}_2)^7.$
\end{itemize}
For the rows $q=3,5,6$, we have that
\[
\Omega_3^{\mathrm{Spin}}(\mathrm{pt})=
\Omega_5^{\mathrm{Spin}}(\mathrm{pt})=
\Omega_6^{\mathrm{Spin}}(\mathrm{pt})=0,
\]
and therefore, the corresponding entries are null.

\subsection{The $E^2$ page}

With all the above, the page $E^2_{p,q}$ for $0\leq p,q\leq 6$ is as shown in table \ref{table2}.\\

\begin{table}[ht]
\centering
\renewcommand{\arraystretch}{1.35}
\begin{tabular}{|l|c|c|c|c|c|c|c|}
\hline
\diagbox{$q$}{$p$} & $0$ & $1$ & $2$ & $3$ & $4$ & $5$ & $6$ \\
\hline
$6$ & $0$ & $0$ & $0$ & $0$ & $0$ & $0$ & $0$\\
\hline
$5$ & $0$ & $0$ & $0$ & $0$ & $0$ & $0$ & $0$\\
\hline
$4$ & $\mathbb{Z}$ & $(\mathbb{Z}_{12})^{2}$ & $\mathbb{Z}_{12}$ & $(\mathbb{Z}_{12})^{3}$ & $(\mathbb{Z}_{12})^{2}$ & $(\mathbb{Z}_{12})^{4}$ & $(\mathbb{Z}_{12})^{3}$\\
\hline
$3$ & $0$ & $0$ & $0$ & $0$ & $0$ & $0$ & $0$\\
\hline
$2$ & $\mathbb{Z}_{2}$ & $(\mathbb{Z}_{2})^{2}$ & $(\mathbb{Z}_{2})^{3}$ & $(\mathbb{Z}_{2})^{4}$ & $(\mathbb{Z}_{2})^{5}$ & $(\mathbb{Z}_{2})^{6}$ & $(\mathbb{Z}_{2})^{7}$\\
\hline
$1$ & $\mathbb{Z}_{2}$ & $(\mathbb{Z}_{2})^{2}$ & $(\mathbb{Z}_{2})^{3}$ & $(\mathbb{Z}_{2})^{4}$ & $(\mathbb{Z}_{2})^{5}$ & $(\mathbb{Z}_{2})^{6}$ & $(\mathbb{Z}_{2})^{7}$\\
\hline
$0$ & $\mathbb{Z}$ & $(\mathbb{Z}_{12})^{2}$ & $\mathbb{Z}_{12}$ & $(\mathbb{Z}_{12})^{3}$ & $(\mathbb{Z}_{12})^{2}$ & $(\mathbb{Z}_{12})^{4}$ & $(\mathbb{Z}_{12})^{3}$\\
\hline
\end{tabular}
\caption{The $E^2_{p,q}$ page in the low-degree range relevant for the computation of $\Omega_6^{\mathrm{Spin}}(BG)$.}
\label{table2}
\end{table}

We see that, on the total diagonal $p+q=6$, only torsion at the primes $2$ and $3$ appears. This property is preserved when passing from one page to the next, since
\[
E^{r+1}_{p,q}
=
\frac{\ker\!\bigl(d_r:E^r_{p,q}\to E^r_{p-r,q+r-1}\bigr)}
{\operatorname{im}\!\bigl(d_r:E^r_{p+r,q-r+1}\to E^r_{p,q}\bigr)}
\]
is always a subquotient of $E^r_{p,q}$.\\

Therefore, all the entries on the diagonal $p+q=6$ in $E^\infty$ still have torsion only at $2$ and $3$. Since moreover, $\Omega_6^{\mathrm{Spin}}(BG)$ is reconstructed from a filtration whose quotients are precisely those terms of $E^\infty$, we conclude that the final group only has torsion at those two primes. This allows us to decompose it as a direct sum of its primary parts:
\begin{equation}
\Omega_6^{\mathrm{Spin}}(BG)\cong \Omega_6^{\mathrm{Spin}}(BG)_{(2)}\oplus \Omega_6^{\mathrm{Spin}}(BG)_{(3)}.
\label{eq:descompprimaria}
\end{equation}

This observation will be the basis of everything that follows because it allows us to study the $2$-primary part and the $3$-primary part separately.

\subsection{Determination of the bordism group}

The explicit determination of the higher pages of the AHSS is difficult without further information about the differentials and possible hidden extensions. We shall therefore follow the approach used in \cite{Debray:2023yrs}, i.e., instead of attempting to determine the full spectral sequence globally, we work prime by prime and replace $B(SL(2,\mathbb Z))$ with a simpler space that has the same localized homological information.\\

For $p=2,3$, define
\[
\Gamma_2=\mathbb Z_4,
\qquad
\Gamma_3=\mathbb Z_3.
\]
The relevant input is that the mappings 
\[
B\Gamma_p\longrightarrow BSL(2,\mathbb Z)
\]
are $p$-local homology equivalences. We shall now explain why the same is true after taking products. Indeed, let
\[
f:X\to X',
\qquad
g:Y\to Y'
\]
be maps inducing isomorphisms in homology with coefficients in $\mathbb Z_{(p)}$. By the K\"unneth theorem with coefficients in $\mathbb Z_{(p)}$, there is a natural short exact sequence
\[
0\to
\bigoplus_{i+j=n}
H_i(X;\mathbb Z_{(p)})\otimes_{\mathbb Z_{(p)}}
H_j(Y;\mathbb Z_{(p)})
\to
H_n(X\times Y;\mathbb Z_{(p)})
\]
\[
\to
\bigoplus_{i+j=n-1}
\operatorname{Tor}^{\mathbb Z_{(p)}}_1
\bigl(
H_i(X;\mathbb Z_{(p)}),
H_j(Y;\mathbb Z_{(p)})
\bigr)
\to 0.
\]
By naturality of the K\"unneth sequence with respect to maps of spaces \cite{Hatcher}, we obtain the following commutative diagram:
\[
\begin{tikzcd}
0 \arrow[r] &
\displaystyle\bigoplus_{i+j=n} H_i(X)\otimes H_j(Y)
\arrow[r] \arrow[d,"\cong"] &
H_n(X\times Y)
\arrow[r] \arrow[d] &
\displaystyle\bigoplus_{i+j=n-1}\operatorname{Tor}(H_i(X),H_j(Y))
\arrow[r] \arrow[d,"\cong"] &
0 \\
0 \arrow[r] &
\displaystyle\bigoplus_{i+j=n} H_i(X')\otimes H_j(Y')
\arrow[r] &
H_n(X'\times Y')
\arrow[r] &
\displaystyle\bigoplus_{i+j=n-1}\operatorname{Tor}(H_i(X'),H_j(Y'))
\arrow[r] &
0 ,
\end{tikzcd}
\]

Since $f$ and $g$ induce isomorphisms on the homology of the factors, they also induce isomorphisms on the corresponding tensor and Tor terms. Therefore, the induced morphism between the two K\"unneth short exact sequences has isomorphisms on the left and right terms. By the five lemma, the middle map
\[
H_n(X\times Y;\mathbb Z_{(p)})
\longrightarrow
H_n(X'\times Y';\mathbb Z_{(p)})
\]
is also an isomorphism for every $n$. Hence $f\times g$ is again a $p$-local homology equivalence.\\

Applying this to the two copies of
\[
B\Gamma_p\longrightarrow BSL(2,\mathbb Z),
\]
we conclude that the product map
\begin{equation}
B(\Gamma_p\times\Gamma_p)
\longrightarrow
B\left(SL(2,\mathbb Z)\times SL(2,\mathbb Z)\right)
\label{eq:productoGamma}
\end{equation}
is also a $p$-local homology equivalence.\\

Using the naturality of the Atiyah--Hirzebruch spectral sequence for $M\mathrm{Spin}$, the map \eqref{eq:productoGamma} induces an isomorphism in $p$-local Spin bordism,
\begin{equation}
\Omega_n^{\mathrm{Spin}}\bigl(B(\Gamma_p\times\Gamma_p)\bigr)_{(p)}
\cong
\Omega_n^{\mathrm{Spin}}\bigl(B(SL(2,\mathbb Z)\times SL(2,\mathbb Z))\bigr)_{(p)}.
\label{eq:isobordismGamma}
\end{equation}
Therefore, the computation of the $p$-primary component can be carried out on the simpler space $B(\Gamma_p\times\Gamma_p)$.\\

We proceed now to compute the primary parts.
Taking $p=3$ in \eqref{eq:isobordismGamma}, we obtain
\[
\Omega_6^{\mathrm{Spin}}\bigl(B(SL(2,\mathbb Z)\times SL(2,\mathbb Z))\bigr)_{(3)}
\cong
\Omega_6^{\mathrm{Spin}}\bigl(B(\mathbb Z_3\times \mathbb Z_3)\bigr)_{(3)}.
\]

Moreover, at odd primes, Spin bordism and oriented bordism coincide, so that
\[
\Omega_6^{\mathrm{Spin}}\bigl(B(\mathbb Z_3\times \mathbb Z_3)\bigr)_{(3)}
\cong
\Omega_6^{SO}\bigl(B(\mathbb Z_3\times \mathbb Z_3)\bigr)_{(3)}.
\]

To compute this last group, we use the Künneth formula for bordism theories due to Peter Landweber~\cite{Landweber}. In this case, for $\ell=3$ and $n=6$, one obtains the exact sequence
\begin{equation}
0 \rightarrow 
\bigoplus_{i+j=6}
\Omega_i^{SO}(B\mathbb Z_3)\otimes
\Omega_j^{SO}(B\mathbb Z_3)
\rightarrow
\Omega_6^{SO}\!\bigl(B(\mathbb Z_3 \times \mathbb Z_3)\bigr)
\rightarrow
\bigoplus_{i+j=5}
\operatorname{Tor}
\Bigl(
\Omega_i^{SO}(B\mathbb Z_3),
\Omega_j^{SO}(B\mathbb Z_3)
\Bigr)
\rightarrow 0.
\label{eq:Landweber3}
\end{equation}

Following \cite{Debray:2023yrs}, in the relevant low degrees, the $3$-local part of $\widetilde{\Omega}^{SO}_*(B\mathbb Z_3)$ appears in odd degrees. In total degree $6$, the contributions come from the pairs
\[
(1,5),\qquad (3,3),\qquad (5,1).
\]
With this, one obtains
\[
\Omega_6^{SO}\bigl(B(\mathbb Z_3\times \mathbb Z_3)\bigr)_{(3)}
\cong
(\mathbb Z_3)^3.
\]

Therefore,

\begin{equation}
\Omega_6^{\mathrm{Spin}}\bigl(B(SL(2,\mathbb Z)\times SL(2,\mathbb Z))\bigr)_{(3)}
\cong
(\mathbb Z_3)^3.
\end{equation}

We now take $p=2$ in \eqref{eq:isobordismGamma}. Then
\[
\Omega_6^{\mathrm{Spin}}\bigl(B(SL(2,\mathbb Z)\times SL(2,\mathbb Z))\bigr)_{(2)}
\cong
\Omega_6^{\mathrm{Spin}}(B\mathbb Z_4\times B\mathbb Z_4)_{(2)}.
\]

At this point, we apply Theorem 10.36 of \cite{Debray:2023yrs}. In degree $6$, this result allows us to identify the $2$-local part of Spin bordism with $ko$-homology:
\[
\Omega_6^{\mathrm{Spin}}(B\mathbb Z_4\times B\mathbb Z_4)_{(2)}
\cong
ko_6(B\mathbb Z_4\times B\mathbb Z_4).
\]

To compute this group, we use a result of \cite{Barcenas}. Theorem 3.2 of that paper gives an additive decomposition of the form
\begin{equation}
ko_*(B\mathbb Z_4\times B\mathbb Z_4)
\cong
\widetilde{ko}_*(B\mathbb Z_4)\oplus
\widetilde{ko}_*(B\mathbb Z_4)\oplus
\widetilde{ko}_*(B\mathbb Z_4\wedge B\mathbb Z_4).
\label{eq:splitko}
\end{equation}

In our case, since $\widetilde{ko}_6(B\mathbb Z_4)=0$, this reduces to
\[
ko_6(B\mathbb Z_4\times B\mathbb Z_4)\cong \widetilde{ko}_6(B\mathbb Z_4\wedge B\mathbb Z_4).
\]

We now use Theorem 5.8 of \cite{Barcenas}, which gives the order of these groups in low degrees. For $n=6$ one has
\[
\log_2\bigl|\widetilde{ko}_6(B\mathbb Z_4\wedge B\mathbb Z_4)\bigr|=5,
\]
from which it follows that
\[
\bigl|\widetilde{ko}_6(B\mathbb Z_4\wedge B\mathbb Z_4)\bigr|=32.
\]
Therefore,
\[
|ko_6(B\mathbb Z_4\times B\mathbb Z_4)|=32,
\]
and hence
\begin{equation}
\left|\Omega_6^{\mathrm{Spin}}\bigl(B(SL(2,\mathbb Z)\times SL(2,\mathbb Z))\bigr)_{(2)}\right|=32.
\end{equation}

Imposing exponent dividing $4$ leaves three candidates,
$\mathbb{Z}_4\oplus\mathbb{Z}_4\oplus\mathbb{Z}_2$,
$\mathbb{Z}_4\oplus(\mathbb{Z}_2)^3$, and $(\mathbb{Z}_2)^5$, of which the generator analysis of Section~\ref{sec:bordism} selects
$\mathbb{Z}_{4}\oplus(\mathbb{Z}_{2})^{3}$, subject to the caveat on hidden
extensions stated there.\\

Combining the 3-primary part with the information obtained for the 2-primary part, we arrive at
\begin{equation}\label{eq:B16}
\Omega^{Spin}_6\big(B(SL(2,\mathbb{Z})\times SL(2,\mathbb{Z}))\big)
\cong \big[\mathbb{Z}_4\oplus(\mathbb{Z}_2)^3\big]\oplus(\mathbb{Z}_3)^3 .
\end{equation}
That is, the odd-primary component is $(\mathbb{Z}_3)^3$ and the $2$-primary component has order $32$, of which the generator analysis of Section~\ref{sec:bordism} selects
$\mathbb{Z}_{4}\oplus(\mathbb{Z}_{2})^{3}$, subject to the caveat on hidden
extensions stated there.

\bibliographystyle{JHEP}
\bibliography{References_z}

\end{document}